\documentclass{article}

\usepackage{graphicx}

\def\abstract#1{\vskip 7mm
        \begin{center}{\large Abstract}\par \smallskip
                \begin{minipage}[c]{12cm}
                        \small #1
                \end{minipage}
        \end{center}
}
\def\title#1{\begin{center}{\Large\bf #1}\end{center}}
\def\author#1{\vskip 5mm \begin{center}{#1}\end{center}}
\def\address#1{\begin{center}{\it #1}\end{center}}
\makeatletter
\@ifundefined{lesssim}{\def\lesssim{\mathrel{\mathpalette\vereq<}}}{}
\@ifundefined{gtrsim}{}{}
\def\vereq#1#2{\lower3pt\vbox{\baselineskip1.5pt \lineskip1.5pt
\ialign{$\m@th#1\hfill##\hfil$\crcr#2\crcr\sim\crcr}}}
\makeatother

\begin{document}

\title{%
Anisotropic scalar field emission from TeV scale black hole
%How does rotating black hole radiate into the brane field?
  %General Relativity and Gravitation
  %\smallskip \\
  %{\large --- Please use this file to complete your manuscript  ---}
}
\author{%
   Daisuke Ida\footnote{E-mail: \texttt{daisuke.ida@gakushuin.ac.jp} }
\address{%
   Department of Physics, Gakushuin University,
   Tokyo 171-8588, Japan }
   Kin-ya Oda\footnote{E-mail: \texttt{odakin@th.physik.uni-bonn.de} }
\address{%
Physikalisches Institut der Universit\"at Bonn, Nussallee 12, Bonn 53115, Germany }
   Seong Chan Park\footnote{E-mail: \texttt{spark@kias.re.kr} }
\address{%
  School of Physics, Korea Institute for Advanced Study (KIAS), Seoul 207-43, Korea%\\
  %207-43 Cheongryangni 2-dong, Dondaemun-gu, Seoul, Korea
}}

 \abstract{
 % Black holes are
 % expected to be produced copiously at future
 % colliders in the scenarios of TeV gravity.
 Black holes are predicted to be copiously produced at
 the CERN Large Hadron Collider in the scenarios of TeV scale gravity.
 We report recent progress in studying
 decay of such a higher dimensional black % hole into Hawking radiation
 hole in $D=4+n$ dimensions through the Hawking radiation
 % in general higher dimensions considering the effects of rotation.
 into brane localized fields
 taking into account its angular momentum
 which is indispensable for realistic simulations.
 % Greybody factors,
 % which determine the emission rate of mass and angular momentum
 % of black hole,
 % are numerically calculated for the brane localized Higgs field
 % in $D=4+n$ dimensions.
 % Interesting new findings include that:
 Presented is the greybody factors for a scalar field emission,
 which confirms our previous results in low energy approximation:
 (i) the existence of super-radiance modes and
 (ii) the non-trivial angular distribution of radiated scalar field.
 % Phenomenological implications of these finding and
 % future directions of studies are briefly sketched.
 Phenomenological implications and future plans are discussed.
 }

\section{Introduction}

%Please write your manuscript in English.\\
%Page limits:  20 pages for the invited speakers, 4 pages for other speakers,
%including poster contributors. \\
%Please download a macro of this form from the workshop web page.
%http://www2.yukawa.kyoto-u.ac.jp/~jgrg14/english/index.html

Black hole is one of the most % mysterious objects in the whole physics.
important key objects in theoretical physics.
% The
Its quantum behavior and % interesting
thermodynamic property have played great roles
% in the progress of theoretical physics
in the path to understand yet unknown quantum theory of gravity
% but the
but a direct experimental % confirmation
test has been believed almost impossible.
Recently, the scenarios of large % or
and warped extra dimension(s)
% suggest a shocking
have led to an amazing possibility of producing black holes
% at future colliders such as
% the Large Hadronic Collider (LHC) of the CERN~\cite{
%Banks:1999gd,Argyres:1998qn,Giddings:2001bu,Dimopoulos:2001hw} and the
at the CERN Large hadron Collider (LHC)
with distinct signals~\cite{Giddings:2001bu,Dimopoulos:2001hw}.
%% NOTE:
%% Writing as above, we can omit Argyres:1998qn and Banks:1999gd.
%% If you want to include, put a foot note as below.
%
%\footnote{
%     See also Refs.~\cite{Argyres:1998qn,Banks:1999gd}
%     for studies before the observation~\cite{Emparan:2000rs}
%     that black holes radiate mainly on the brane.}
%
%
% Main purpose of this talk is to discuss the decay signals of such black
% holes into Hawking radiation of brane field~\cite{Hawking:1974sw}.
% Unfortunately, we could not perform any kind of
% trustful calculation in the domain of Planckian energy where the
% quantum effects of gravity could not be neglected.
% However once the mass of black hole is much larger than the Planck scale,
% the semi-classical approximation would still work quite
% well~\cite{'tHooft:1987rb} and we can obtain greybody factor for
% massive black hole in the Trans-Planckian energy domain where the
% size of event horizon were to be much larger than the Planck
% length and one could safely neglect the stringy effects
%~\cite{IOP-I,Kanti:2002ge}.
When the center-of-mass energy of a collision exceeds the Planck scale,
which is of the order of TeV here, the cross section is
dominated by a black hole production~\cite{'tHooft:1987rb}.
%% NOTE:
%% I (KO) think that above citation is unnecessary
%% but leave it if both of you want it.
In this trans-Planckian energy domain,
the larger the center-of-mass energy is,
the larger the mass of the resulting black hole,
and hence the better its decay
is treated semi-classically via Hawking radiation.
Main purpose of this talk is to discuss such decay signals.

In previous publications, we have pointed out that
the production cross section of a black hole \emph{increases} with its
angular momentum, so that the produced
black holes are highly rotating~\cite{IOP-I,Proceedings}.
%% NOTE:
%% Above statement is rather sloppy but
%% the point is easier understood this way, I (KO) think.
(See also Ref.~\cite{Park:2001xc} for an earlier attempt.)
% Actually
The form factor for the production cross section,
% estimated in the paper~\cite{IOP-I},
% is quite in good agreement with the results of
% other studies~\cite{Eardley:2002re,Yoshino:2002tx}
taking this rotation into account~\cite{IOP-I},
is larger than unity and increases with the number of extra dimensions $n$.
(The result is in good agreement with
an independent numerical simulation of a classical gravitational collision of
two massless point particles~\cite{Yoshino:2002tx}.)
% also see Ref.~\cite{Yoshino:this}.
% This result that
% the size of the form factor % would increase
% increases with the % larger
% number of extra dimensions,
% when we correctly include the effect of rotation
% and this tendency
We note that this form factor is hardly interpretative
% is hardly understandable without consideration of
without considering the angular momentum.
% With this respect, we have claimed that
It is essential to take into account the angular momentum of the black hole
%when we calculate the greybody factor of brane field emission~\cite{IOP-I}.
when we perform a realistic calculation of its production and evaporation.

Black holes radiate mainly into the standard model fields
that are localized on the brane~\cite{Emparan:2000rs}.
The master equation for brane field perturbation on
the Myers-Perry higher dimensional rotating black hole % metric~\cite{Myers:1986un}
background has been derived~\cite{IOP-I} for general fields with spin
zero, one half and one (that is, for all the standard model fields).
% and it will be used in our present and future projects.
In the paper~\cite{IOP-I},
we have shown %the
analytic expressions for the greybody
factor of $D=5$ (Randall-Sundrum) black hole by solving the master
equation under the low energy approximation of radiating field.
In this talk % we would like to generalize the result by considering
we present a generalized result to % general
higher dimensional black hole in $D >5 $ for brane localized
scalar field without relying on the low energy approximation and
discuss
% interesting physics with
% non-zero angular momentum such as super-radiance mode.
its physical implications.

\section{Master equation for Brane field perturbation} % ($D=4+n$)}
The Teukolsky equation for spin $s$ brane field %~\cite{Teukolsky:1972my}
in $D=4+n$ dimensional Myers-Perry black hole background has been
derived and the resultant
equation is shown to be separable into the angular and the radial
parts in our previous paper~\cite{IOP-I}.
The angular function $S$,
% (so called `` the spin weighted spheroidal harmonics'')
the spin weighted spheroidal harmonics,
is a regular function on
$[0,\pi]$ which satisfies the equation
\begin{eqnarray}
  0={1\over\sin\vartheta}{d\over d\vartheta}\left(\sin\vartheta{dS\over
  d\vartheta}\right)
   +\left[(s-a\omega\cos\vartheta)^2
   -\left({s\over\tan\vartheta}+{m\over \sin\vartheta}\right)^2+E-2s^2\right]S,
\end{eqnarray}
% for eigenfunction ${}_s E_{lm}$.
with the eigenvalue $E$ for a given angular mode $(l,m)$.
This angular equation is exactly
the same as the usual four dimensional Kerr black hole.
The radial function % $R_s(r)$
$R$ satisfies the equation
\begin{eqnarray}
0=\Delta^{-s} \frac{d}{dr}\left(\Delta^{s+1}\frac{d R}{dr}\right)
+\left[{K^2-isK\Delta,_r\over\Delta}-\lambda_s+4isr\omega+s(\Delta,_{rr}-2)\right]R, \label{starting_eq}
  \end{eqnarray}
 where
  $\Delta = r^2+a^2-(r_H^2+a^2)\left(r/r_H\right)^{-(n-1)}$,
  $K = (r^2+a^2)\omega-ma$  and
  $\lambda_s = E-s(s+1)-2ma\omega+(a\omega)^2$
  with $r_H$ and $a$ being the horizon radius and the rescaled dimensionless angular momentum
  of the black hole. The $\omega$ is the frequency of the radiated field
that was assumed to be small in the derivation of the greybody factors in Ref.~\cite{IOP-I}.
If the number of extra dimensions $n$ is put to be zero,
the radial equation % becomes
reduces to the usual Teukolsky equation in four dimensions.

%In the paper~\cite{IOP-II}, we
%will further clarify the physical % meaning
%implications of the master equation
%by considering the cases with the
%negative $s$: $-1/2$ and $-1$ and will show the details of derivation.
%% NOTE:
%% Are you both sure about what to do for negative $s$?
%% I would suggest to remove it but if both of you are sure, please leave it.
%%% I (DI) agree with K on this point

\section{Greybody factor for Brane scalar field emission} % ($D=4+n$)}
Evolution of the mass and the angular momentum of a black hole can be
determined by emissions via Hawking radiation
% law which is valid in semi-classical energy domain:
up to the final Planck phase where a semi-classical description breaks down
\begin{eqnarray}
-\frac{d}{dt}\left(
\begin{array}{c}
  M \\
  J
\end{array}
\right)=\frac{1}{2\pi} \sum_{s,l,m}g_s \int d\omega
\frac{\Gamma_{s,l,m}(\omega)}{e^{\omega- m\Omega/T}\mp 1}
\left(\begin{array}{c}
  \omega \\
  m
\end{array}
\right),\label{MJeq}
\end{eqnarray}
where $g_s$ is the effective degrees of freedom for spin $s$
field, $\Omega$ is the surface angular velocity of the rotating hole,
$T$ is Hawking Temperature and $\Gamma_{s,l,m}$ is the greybody
factor for a given angular mode $(l,m)$. The analytic expression for
$\Gamma_{s,l,m}$ for five dimensional black hole has been % already
obtained in % the paper
Ref.~\cite{IOP-I}.
%Here we present a result for brane scalar
%emission with $D=10$ bulk dimensions in Fig.~1.

One immediate observation is that the Hawking radiation of
$m>0$ mode dominates over that of $m<0$ mode when $\Omega \rightarrow \infty$,
i.e.\ when the hole is highly rotating.
% This is extremely interesting for super-radiance mode with the negative
% greybody factor $\Gamma_{s,l,m}< 0$ which are obtained in
% higher dimensional black hole for the modes with $m>0$.
In this limit, the modes with $m>0$ show super-radiance with negative
greybody factor $\Gamma_{s,l,m}<0$, namely Hawking \emph{absorption}.
% Thus it could be generally possible that the black hole could actually
% {\it increase} in its size when the hole is highly rotating!
This can in principle result in the initial \emph{increase} of the area
and entropy for a highly rotating black hole~\cite{Page:1976ki}.

%The tasks we did in this project could be schematically summarized
%as follows:
%The greybody factors are obtained as follows:
%\begin{enumerate}
%    \item % First, specify
%%    Put a purely ingoing boundary condition
%%    at near horizon region by extracting out the outgoing wave
%%    contamination, which is especially severe for $s=1$
%%    since the outgoing wave grows much faster than the ingoing wave.
%     Put the purely ingoing boundary condition at the horizon.
%    %% NOTE:
%    %% Please check out and in are correct.
%    %% Do we need this to say about contamination here? I (KO) would say no.
%    \item % Second,
%    Numerically integrate the master equation
%    from the near horizon to infinity.
%    \item % Third,
%    Separate the numerical result into the ingoing
%    and outgoing parts at infinity. The ratio of the numerical
%    coefficients of the ingoing ($Z_{\rm in}$) and outgoing ($Z_{\rm out}$)
%    solutions determines the absorption cross-section of the given mode
%    ($l,m$).
%    \item
%    Then we can read out the greybody factor immediately:
%\begin{eqnarray}
% \Gamma_{0, l, m}=\left[\frac{dE_{\rm out}/dt}{dE_{\rm
%in}/dt}\right]_{\rm scalar}= \frac{|Z_{\rm out}|^2}{|Z_{\rm
%in}|^2}.
%\end{eqnarray}
%\end{enumerate}
%% NOTE:
%% Do we need this (above) paragraph at all?
%% I (KO) would say no.
%% I would prefer removing the whole paragraph.
%% I(DI) also feel it redundant about this paragraph.

The numerical results of the greybody factors
for the $D=5$ black hole is in good agreement with the analytic
expression in~\cite{IOP-I}.
As an example for $D>5$ case, we % would just
present % the
a result of $l=1, m=1$ mode, where the effects of rotation and the
existence of super-radiance (negative greybody factor) can be
clearly seen. Complete results for $D\geq 4$ will be shown
in~\cite{IOP-II}. We plot for a $D=10$ non-rotating hole $a=0$
(the thin line) and for rotating one $a=1$ (the thick line) in
Figure~1.
% We can observe a super-radiance in $a=1.0$ case.
A super-radiance does not exist in negative mode $m=-1$
which is not presented here.
% However, we do not observe
% super-radiance mode with the negative value of $m$ which is not
% presented here.
When black hole is highly rotating, i.e.\ having large $a$,
the difference between $m=1$ and $m=-1$ modes leads to
the non-trivial angular distribution of
Hawking radiation even for a scalar field emission.
This % could be one of the
is a clear signature of the {\it rotating} black hole.
According to the analytic expression for
the $D=5$ dimensional black hole~\cite{IOP-I},
the criteria of super-radiance in $l=1$ mode is: $ \omega - m \Omega
< 0$ since $\Gamma \propto \omega - m \Omega$. The numerical
result is % quite
consistent with this: for $a = 1.0$ which is
denoted as the thick line in the figure, we obtain the
super-radiance mode when $\omega \lesssim \Omega$.
%\footnote{We use the unit where the size of black hole is fixed $r_H =1$.}
Further results of the general $D=4+n$ dimensional black holes
with varying angular momentum for the general angular modes $(l,m)$ will be
presented elsewhere~\cite{IOP-II}. % in the else place~\cite{IOP-II}.

% D=5 , l=1, m=1,0,-1
%\begin{figure}[thb]
 % % Requires \usepackage{graphicx}
  %\includegraphics[scale=0.3]{5D011.eps} \\
  %\includegraphics[scale=0.3]{5D0lm1.eps}\\
  %\caption{$5D,l=m=1$}%\label{10D,l=m=1}
%\end{figure}

%\begin{figure}[thb]
%  \includegraphics[scale=0.3]{5D01m1.eps}\\
 % \caption{$5D,l=1,m=-1$}%\label{10D,l=m=1}
%\end{figure}

%\begin{figure}[thb]
%  \includegraphics[scale=0.3]{5D010.eps}\\
 % \caption{$5D,l=1,m=0$}%\label{10D,l=m=1}
%\end{figure}

% D=10, l=1, m=1,0,-1

%\begin{figure}[thb]
 % \includegraphics[scale=0.3]{10D010.eps}\\
  %\caption{$10D,l=1,m=0$ }%\label{10D,l=1,m=0}
%\end{figure}

%\begin{figure}[thb]
 %\includegraphics[scale=0.8]{10D011.eps}\\
%\caption{Greybody factor for $D=10,l=1,m=1$ mode. Red line denotes
%$a=0$ non-rotating case, pink line denotes $a=0.9$ and green and
%bluish green line denote $a=0.3$ and $a=0.6$, respectively.
%Super-radiance modes in the small energy region are clearly observed. }%\label{10D,l=m=1}
%\end{figure}

%\begin{figure}[thb]
  %\includegraphics[scale=1]{10D01m1.eps}\\
  %\caption{$10D,l=1,m=-1$}%\label{10D,l=1,m=-1}
%\end{figure}

\begin{figure}[thb]
 \includegraphics[scale=0.8]{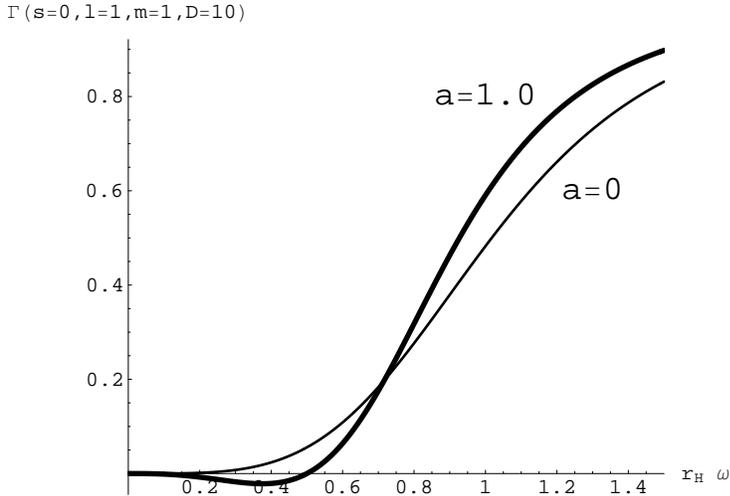}\\
\caption{
   Greybody factor for $D=10$, $(l,m)=(1,1)$ mode.
   The thin and thick curves
   denote non-rotating $a=0$ and rotating $a=1.0$,
   respectively.
   Super-radiance % modes
   in the small energy region can be observed for the latter.}%\label{10D,l=m=1}
\end{figure}

\section{Discussion}
Here we have explained the importance of the angular momentum
when one considers TeV scale black hole production and evaporation.
%A new numerical result is reported: % the greybody factor for
New numerical results are reported:
greybody factor for the brane scalar emission
from a general $D=4+n$ dimensional rotating black hole
without relying on the low frequency expansions.
% Numerical integration has been done from the purely ingoing boundary
% condition at the near horizon region to infinity where the wave
% function can be separated into the ingoing wave and the outgoing
% wave. From the ratio between the ingoing and outgoing energy
% fluxes, we read out the greybody factor for each mode of
% emission.
% The results are obtained for general $D\geq 4$
% dimensional cases and the various angular modes $(l,m)$.
We observe super-radiance in $m>0$ modes for $\Omega>0$.
The existence of super-radiance mode is crucially important
new feature of the rotating black hole.
Nontrivial angular dependence of the
emission is observed even for the scalar field, which will be
a signature of {\it rotating} black hole.

To understand the actual evolution of a black hole and
to predict the collider signature, % there
we need further investigations.
% Some words on future directions are given as follows. First,
It is important to determine the greybody factors for spinor and
vector fields \cite{IOP-I, Kanti}. The evolution of the angular
momentum and the mass is determined by integrating
Eq.~(\ref{MJeq}), once all the greybody factors are determined.
In particular, the % time scale of
spin-down phase, which has been simply neglected so far,
can be precisely described. % determined.
One can in principle determine the angular momentum of the produced black hole
from the nontrivial angular distribution of the signals.
The super-radiance mode is a unique feature of rotating black hole and
its phenomenological implications should be clarified.
Detailed simulations of black hole evaporation at LHC are to be done. %  by the correct greybody factor.
% on which we already have
% obtained some progresses and will be presented soon.
Some of these points will be presented soon~\cite{IOP-II}.

%%%%%%%%%%%%%%%%%%%%%%%%%%%%%%%%%%%%%%%%%%%%%%%%%%%%%%%%%%%%%%%%%%%%%%%
%\section{Deadline: January 21, 2005}
%
%  Please prepare a postscript or PDF file of your manuscript. \\
%We ask you to use a systematic file name (your presentation
%      number).ps or (your presentation number).pdf. Your presentation
%      number (PRE-XX or POS-XX) will be found at:%
%
%http://www2.yukawa.kyoto-u.ac.jp/~jgrg14/english/presen\_list.html%
%
%\begin{enumerate}
%\item e-mail to

%jgrg14@yukawa.kyoto-u.ac.jp

%with subject including jgrg14\_proc (your full name) (your
%presentation number)

%\begin{verbatim}
% ex.) Subject: jgrg14_proc Takahiro Tanaka PRE-99
%\end{verbatim}

       %when your file size is large, please COMPRESS it.

%\item postal mail to
%\begin{verbatim}
%  JGRG14 (T. Tanaka)
%  Theoretical Astrophysics Group,
%  Department of Physics, Kyoto University,
%  Kyoto 606-8502, Japan
%\end{verbatim}
%\end{enumerate}
%%%%%%%%%%%%%%%%%%%%%%%%%%%%%%%%%%%%%%%%%%%%%%%%%%%%%%%%%%%%%%%%%%%%%%%%%%

\end{document}